\documentstyle[aps,prl,twocolumn]{revtex}

\begin{document}
\draft
\title{A New Transport Regime in the Quantum Hall Effect}
\author{D. Shahar, M. Hilke, C.C. Li and D. C. Tsui}
\address{Department of Electrical Engineering, Princeton University,
Princeton New Jersey, 08544}
\author{S. L. Sondhi}
\address{Department of Physics, Princeton University,
Princeton New Jersey, 08544}
\author{M. Razeghi}
\address{Department of Electrical Engineering,
Northwestern University, Evanston Illinois, 60208}

\date{\today}
\maketitle

\newpage {\bf Our evolving understanding of the dramatic features of
charge-transport in the quantum Hall (QH) regime has its roots in the
more general problem of the metal-insulator transition. Conversely, the
set of conductivity transitions observed in the QH regime provides a fertile
experimental ground for studying many aspects of the metal-insulator
transition. While earlier works\cite{wei,Koch91} tend 
to concentrate on transitions
between adjacent QH liquid states, more recent works 
\cite{Jiang93,TWang94,Hughes94,Alphenaar:2terminalHI,Shahar:Univ,Okamoto:Si}
focus on the transition from the
last QH state to the high-magnetic-field insulator.
Here we report on measurements that
identified a novel transport regime which is distinct from both,
the fully developed QH liquid,
and the critical scaling regime believed to exist
asymptotically close to the transition at very low temperatures ($T$'s).
This new regime appears to hold in a wide variety
of samples and over a large range of magnetic field ($B$) and
temperature. It is characterized by a remarkably simple
phenomenological scaling of the longitudinal resistivity
($\rho_{xx}$), which is the center of this letter, and is not
understood theoretically.}

We begin by focusing on a recent set of
observations that directly begot some of the results presented here.
In Refs. \cite{Shahar:dualitySC,Shahar:natureHI}, the
observation of a new symmetry was reported, relating the transport
properties of the QH liquid to those of the adjacent insulator. For
the case of the $\nu=1$ to insulator transition, this
symmetry is summarized by:
\begin{equation}
 	\rho_{xx}(\Delta \nu)=1/\rho_{xx}(-\Delta \nu)
 	\label{rxxDuality}
\end{equation}
where $\nu$ is the Landau level filling factor,
$\Delta \nu= \nu-\nu_{c}$ and $\nu_{c}$ is the critical
$\nu$ of the transition (see the inset of Fig. 1 for the
identification of $\nu_{c}$).
Remarkably, a similar symmetry holds at transitions from the $1/3$
fractional quantum Hall (FQH) state to the insulator, if one replaces
the $\nu$'s in Eq. \ref{rxxDuality} with those of
composite fermions\cite{JKJain89}.
In addition, in ref. \cite{Shahar:dualitySC} we showed
that a generalized relation
holds, within experimental error, even for the non-linear regime of
transport, and suggested the possibility that duality symmetry
underlies this relation\cite{Shahar:dualitySC,Shimshoni97}. More
recently, a similar relation was observed in Si-MOSFET samples near the
$B=0$ conductor-insulator transition\cite{Simonian:reflection}, which 
raises the question whether
a more general explanation may exist for the symmetry\cite{Sudip:exp}.

In the remainder of this paper we shall present and discuss a
view of the $\rho_{xx}$
data where the symmetry of Eq. \ref{rxxDuality} is a
straightforward ingredient. We begin by plotting,
in Fig. 1, $\rho_{xx}$ vs. $\nu$ for a low
mobility ($\mu=30000$ cm$^{2}$/Vsec), low density ($n=3\cdot
10^{10}$ cm$^{-2}$), InGaAs/InP sample,
in the range of $0.4<\nu<0.8$ which
includes the $\nu=1$-to-insulator transition ($\nu=0.562$),  at
several $T$'s between 0.072 and 2.21 K. Rather than plotting the data using
the conventional linear ordinate (see inset of Fig. 1),
we chose a log scale, which clearly reveals
a distinct $\nu$ dependence of $\rho_{xx}$:
\begin{equation}
  	\rho_{xx}=e^{\frac{-\Delta \nu}{\nu_{0}(T)}}
  	\label{explaw}
\end{equation}
where $\rho_{xx}$ is measured in units of
its critical value, $\rho_{xxc}$ ($=29.6$ k$\Omega$ for this sample),
a normalization which we adopt throughout this letter, and
$\nu_{0}(T)$ is a $T$-dependent logarithmic slope,
introduced here for the first time.
Eq. \ref{explaw} leads to the first
new result of this work: Evidently, data described by it
has the necessary symmetry required by Eq. \ref{rxxDuality}.
The range over which Eq. \ref{explaw} holds is quite extensive:
As can be seen in Fig. 1, Eq. \ref{explaw}
is a good description of our data for more than 4 orders
of magnitude in $\rho_{xx}$ with only small,
non-systematic, deviations over a large range of $\nu$.
At lower $\nu$'s systematic deviations appear due to nonohmic
effects in the insulator while in the QH phase
our measurement is effectively limited
to $\nu<0.8$ where $\rho_{xx}$ becomes too small to detect at low $T$.
Therefore, we can only put a lower bound on the $\nu$-range of the
applicability of Eq. \ref{explaw}.

So far, we demonstrated that $\rho_{xx}$ data near the  QH-insulator
transition can be described by a single, well-defined, expression
that holds on both sides of the transition. This
is potentially useful from the prevalent theoretical standpoint, which
asserts that transport coefficients such as $\rho_{xx}$ should be
described, near the transition, by a 
scaling form\cite{Huckestein:RMP,Sondhi:RMP}:
\begin{equation}
	\rho_{xx}=\rho_{xxc}f(\frac{\Delta\nu}{T^{1/z\nu}})
	\label{scaling}
\end{equation}
where $f(X)$ is a universal function, and $\nu$, $z$ and $\rho_{xxc}$
are the critical exponents and amplitude of the transition,
respectively, and are also expected to be universal\cite{KLZ} ($\nu$
here should not be confused with the LL filling factor).

It is therefore natural to try and associate the scaling function,
$f(X)$, with the experimentally derived Eq. \ref{explaw}.
This could lead to a determination of the product of the scaling
exponents, $z\nu$, which can then be compared with theoretical
estimates\cite{Huckestein:RMP}. We will next
show that rather surprisingly, upon trying this association
using the experimental data, we are faced with a contradiction
that leads us to conclude that our data cannot be described by
a scaling form. Instead, a form of behavior new to two-dimensional 
electron systems (2DES) 
at high $B$ emerges, for which no theoretical understanding yet exists.

Inspecting Eqs. \ref{explaw} and \ref{scaling} shows that the
simplest way of comparing them is by fitting the $\nu_{0}(T)$ data to
$T^{1/z\nu}$. In fig. 2a we attempt this comparison by plotting
the $\nu_{0}$ values obtained from the sample of Fig. 1 vs. their
$T$'s, using a log-log graph. Clearly, the fit (dashed line, with the optimal 
value of $1/z\nu=0.64$) fails
at low $T$, and other attempts using a modified $1/z\nu$ power are equally
unsuccessful, as they should be: The data significantly
deviates from a power-law behavior expected from the scaling
prediction (a straight line on a log-log plot).

An alternative view is revealed in Fig. 2b, where
the data are plotted on a linear scale: They are suggestive
of a linear dependence,
\begin{equation}
	\nu_{0}(T)=\alpha T + \beta ,
	\label{linearT}
\end{equation}
with $\alpha=0.088$ K$^{-1}$ and $\beta=0.0537$ for this sample. It
is the offset, $\beta$, rather than the linear dependence itself,
which renders the scaling description and its associated power-law
$T$ dependence unsuitable to describe our data. (Even if one does not
accept the linear description of the data it is still clear, unless a
different conduction mechanism takes over at still lower $T$'s, that
$\nu_{0}(T=0)\neq 0$ and the transition is of finite width.)
We emphasize that this linear dependence, with $\beta>0$,
is a very general result seen in the 20 
samples studied, which are made from various semiconductor
materials and fabrication techniques.
The parameters
$\alpha$ and $\beta$ are sample-dependent, with $\alpha=0.034-0.24$ K$^{-1}$
and $\beta=0.003-0.054$. We found that their variations are
correlated and $\beta/\alpha$, which defines a new $T$ scale for the
conduction process, seems to be close to $0.5$ K for InGaAs/InP
samples and $50$ mK for GaAs/AlGaAs samples. The physical significance
of this temperature scale is yet unclear.

Putting Eqs. \ref{explaw} and \ref{linearT} together leads
to our final phenomenological form:
\begin{equation}
  	\rho_{xx}=e^{(\frac{-\Delta\nu}{\alpha T + \beta})}.
	\label{rlaw}
\end{equation}
This compact expression describes the $\rho_{xx}$ data over a very
large range of both $T$ and $\nu$, and holds for all our samples.

Since our failure to `scale' our data in the usual way is a direct
result of the finite $\beta$, it becomes
imperative to establish whether this observation is an
integral part of the transition, or an experimental artifact,
with $\beta$ vanishing in the appropriate limits of infinite sample size
and zero excitation current.
Naturally, a definite answer to any of these questions is, at best,
hard to obtain.
Nevertheless, we would like to present several compelling arguments
that alleviate some of these concerns. First, all our samples are
relatively large, ranging from $100\times 300$ $\mu$m to
$1\times 1$ mm in size. Even at mK temperatures, it is not expected
that such samples will exhibit finite-size effects.
Second, we found that our results, namely the parameters presented in
Table 1, are independent (within error) of excitation currents that
are as much as 10 times larger than the currents used in this study
($0.01-1$ nA).
Thus, non-ohmic heating effects are unlikely to be the cause of the
finite $\beta$. Increasing the current further
resulted in significant deviation from the behavior described by Eq.
\ref{explaw}.
Third, the results presented in this letter are quite
general to all our samples, which rules out the possible that
gross sample inhomogeneities and imperfections (on a scale comparable 
to the sample size) dominate the transport. Smaller scale 
inhomogeneities and disorder are certain to play a role, which is not 
yet understood.

The next logical step in light of the simple and general form
$\rho_{xx}$ takes is to
consider the behavior of the Hall resistivity, $\rho_{xy}$.
Unfortunately at this stage a clear experimental answer can not be
given, due to the lack of consistent results. For some of our
samples we have, however, obtained a $\rho_{xy}$ value that
remains quantized beyond the QH state into the insulating phase
\cite{Shahar:natureHI}.
Incidentally, given the symmetry of $\rho_{xx}$, a $\rho_{xy}$ 
which is indeed quantized on both sides of 
the transition is a prerequisite for a symmetric $\sigma_{xx}$.

The inconsistency of our data with a scaling description is
troubling if we recall several earlier results that were taken to
indicate the proximity of the transition region to a 
quantum critical point\cite{Sondhi:RMP}. While two of these
results, namely the existence of a universal (within 25\%) critical
resistivity at the transition\cite{Shahar:Univ,Wong95}
and the observation of a reflection
symmetry consistent with charge-flux duality\cite{Shahar:dualitySC}
are consistent with the data presented here, the indication
of scaling behavior observed by Wei et al.\cite{wei} is clearly not.
This is particularly puzzling
since some of our measurements were done on InGaAs/InP samples that
are from the same growth as the sample in Ref. \cite{wei}. We can
not rule out a scenario in which the sample of Ref. \cite{wei} is of
exceptional homogeneity and is therefore a better representative of
the ideal theoretical case. It is conceivable that 
our samples would crossover to similar scaling behavior at (yet 
inaccessible) lower $T$'s. We may remark, however, that a universal critical
amplitude was not observed for the transitions studied by Wei et al., 
which is actually in conflict with the scaling framework. We should also 
note that our previous scaling analyses of the QH-insulator 
transition\cite{Pan:scaling}, which attempted to 
``collapse'' a narrow range of $\rho_{xx}(\nu,T)$ data near the transition,
were fairly successful despite the fact that the data follows Eq. 
\ref{rlaw} with $\beta>0$. Attempts to collapse numerical ``data''
generated using Eq. \ref{rlaw} with parameters obtained 
from our samples acutely demonstrate that using a limited range 
of data can give false indication of quantum-critical behavior.
Needless to say, they also call into question our previous belief that the
symmetries of the transport reflect symmetries of an underlying 
quantum-critical point.

In this letter we argued that the $\nu$ range where Eq.
\ref{rlaw} holds is not a critical region of a quantum phase
transition. It is important to note that this regime is distinct
from a fully developed QH liquid or an insulator as well, for it extrapolates,
through the positive-definite $\beta$, to a finite resistivity at $T=0$ 
for $\nu$ both bigger and smaller than $\nu_{c}$. 
Whether this new regime of 2DES at high-$B$ can be sharply defined
is unclear. But even if it were to give way to one,
or more, of the familiar phases of 2DES
in the ideal limit of vanishing
$T$'s and highly homogeneous material, the extremely wide range of
applicability of Eq. \ref{rlaw} both in $T$ and $\nu$, over such a
diversity of samples, coupled with its introduction of a new, 
$B$-dependent temperature scale, $\beta/\alpha$, 
to the conduction process render
this new regime a very important, as well as interesting, topic of 
research.

Correspondence and requests for materials should be addressed to  D.S. 
(e-mail: dannys@princeton.edu).

\begin{figure}
\caption{$\rho_{xx}$ vs. $\nu$ at $T=72$, 293, 518, 940, 
1461 and 2210 mK, for an InGaAs/InP sample, RA609C.
Inset: $\rho_{xx}$ (linear scale) vs. $\nu$ at $T=293$, 940, 1461  
and 2210 mK. The arrow indicates the transition at $\nu_{c}=0.562$, 
which is the common crossing point of the $\rho_{xx}$ isotherms.
}
\end{figure}

\begin{figure}
\caption{(a) A log-log graph of $\nu_{0}$ of Eq. \ref{explaw} vs $T$
obtained from a fit of the $\rho_{xx}$ traces of sample RA609C (Fig. 1) 
to Eq. \ref{explaw}. (b) Same as (a), plotted using a linear graph. 
Solid lines are least-square fit to Eq. \ref{linearT} resulting 
in $\alpha=0.0878$ K$^{-1}$ and $\beta=0.05367$. Dashed line in (a) is 
the optimal power-law fit: $\nu_{0}\sim T^{1/z\nu}$, with $1/z\nu=0.64$.
Inset: Same as (b), for a GaAs/AlGaAs sample, with the linear fit (solid 
line) resulting in $\alpha=0.24$ K$^{-1}$ and $\beta=0.014$.
}
\end{figure}

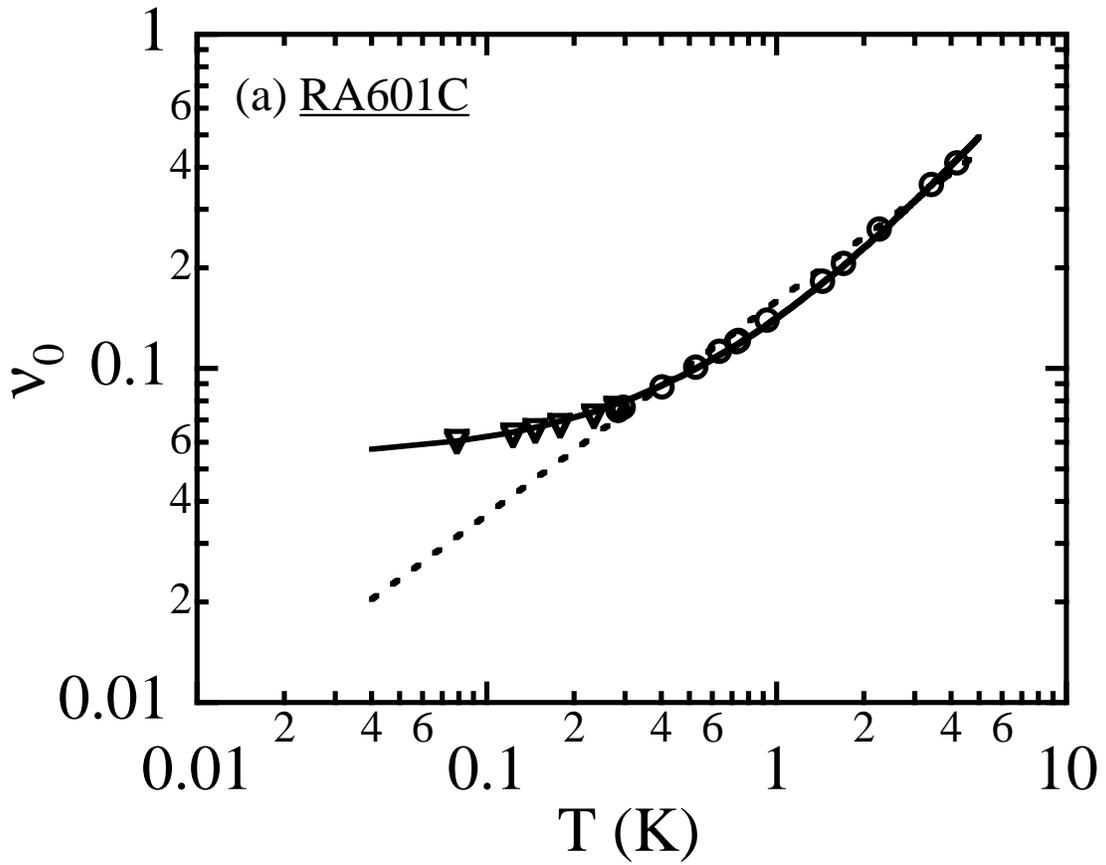
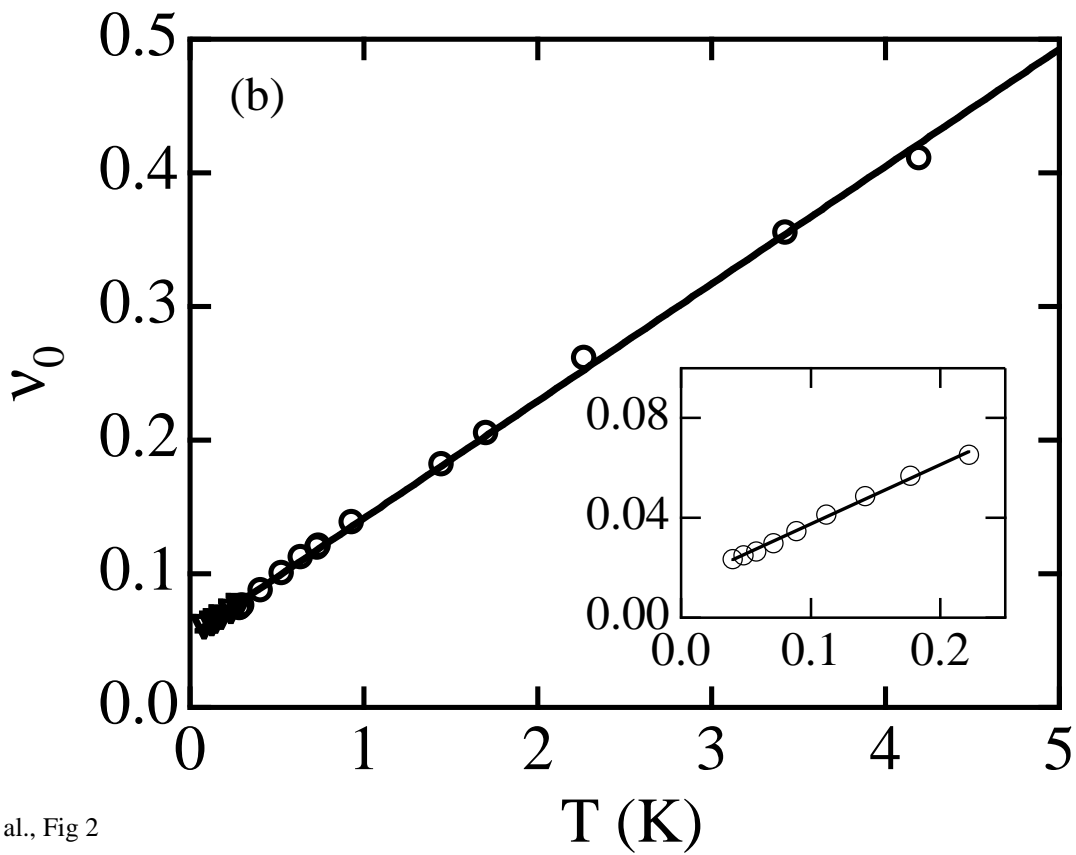
\end{document}